\documentclass{aa}
\input psfig.sty

\def\Msun{$M_\odot$}
\def\Teff{$T_{\rm eff}$}
\def\Mdot{$\dot{M}$}
\def\Mbol{$M_{\rm bol}$}
\def\Tbce{$T_{\rm bce}$}
\def\tauf{$\tau_F$}
\def\simgt{\lower.5ex\hbox{$\; \buildrel > \over \sim \;$}}
\def\simlt{\lower.5ex\hbox{$\; \buildrel < \over \sim \;$}}

\begin{document}

  \thesaurus{06 
	     ( 08.16.4;  
	       08.05.3;  
	       08.13.2)} 
   \title{Lithium and mass loss in massive AGB stars in the Large Magellanic
   Cloud.}

   \author{P. Ventura \inst{1} F. D'Antona \inst{1} \and
	   I. Mazzitelli \inst{2}}

\offprints{P. Ventura}

   \institute{Osservatorio Astronomico di Roma,
	      Monte Porzio Catone (Rome), I-00040, Italy \\
	      email: paolo@coma.mporzio.astro.it
	      \and
	      Istituto di Astrofisica Spaziale C.N.R., Via Fosso del 
	      Cavaliere, I-00133 Rome, Italy \\
	      email: aton@hyperion.ias.rm.cnr.it
	      }

 \date{Received 21 April 2000; accepted ???????}

   \titlerunning{Lithium production in LMC stars}

   \maketitle

\begin{abstract}

The aim of this work is to use full evolutionary models to
derive observational constraints on the mass loss rate of the
upper Asymptotic Giant Branch (AGB) stars.
The observations used to constrain the models are: i) the relative number of
luminous Lithium rich AGBs in the
Magellanic Clouds, with respect to the total number
of AGBs populating the luminosity range $-6 \geq$ \Mbol $ \geq -7$; ii) the
s--process enhancement of the same sample. The calibration of the mass loss
rate we obtain
gives feedbacks on the interpretation of observational data of obscured AGBs,
and allows us to provide consistent lithium yields for these stars, to be
used to constrain the galactic chemical evolution.

We find that:
a) we can put lower and upper limits to the mass loss rate during the AGB
phase;
b) after a ``visible" phase, the models evolve into a phase of strong mass
loss, which can be identified with the obscured OH/IR stars accessible
only in the infrared; the models well reproduce the Period--\Mbol\
loci of the obscured AGBs (Wood et al. 1992);
c) the most massive AGBs (mass of progenitors, hereinafter $M_{ZAMS}$, 
$\sim 6M_{\odot}$) are extremely luminous
(\Mbol $\sim -7.2$ to $-7.5$); d) The
lithium yield increases with the mass loss rate and with the total stellar
mass, being maximum for AGB stars close to the lower limit for carbon
semi-degenerate ignition. However, the mass loss calibration obtained in this
work implies that massive AGBs {\it do not} contribute significantly to the
lithium enrichment of the interstellar medium.


\end{abstract}

\section{Introduction}

Following the pioniering studies by Schwarzschild \& Harm (1965, 1967) and
Weigert (1966), it is now a quarter century since the first extensive
modeling of Asymptotic Giant Branch (AGB) structures, in which, above the
carbon oxygen core, hydrogen and helium alternatively burn in shells, with
the He-burning phase being initiated by a thermonuclear runaway (Thermal
Pulse --TP-- phase, e.g. Iben 1981). The convective shells developped during
the TPs, and the following ``dredge up" of inner nuclearly processed
material to the surface (Iben 1975) leads to the formation of Carbon and
s--process enhanced stars. Already two decades ago (see Iben 1981) 
it was realized that
in the LMC, where the AGB luminosities are more reliable than in the
Galaxy, there were no Carbon stars more luminous than \Mbol $\sim -6$
(Blanco et al. 1980), contrary to the theoretical expectations on the
extension of the AGB up to \Mbol $\simgt -7$. This old finding is confirmed
by recent studies (Costa \& Frogel 1996). 

Later on, very luminous AGBs
were discovered (Wood et al. 1983): they were oxygen rich
(M--type) stars, and were few compared to the numbers expected if massive
AGBs evolve at the theoretical nuclear rate of $\sim 1$mag$/10^6$yr.
This is confirmed also by the scarciness of these stars in the MCs clusters
(e.g. Frogel et al. 1980, Mould \& Reid 1987). \\
It is generally accepted that the lack of a C--star stage above \Mbol $\sim
-6$\ is due to nuclear processing at the bottom of the convective mantle of
massive AGBs (Hot Bottom Burning, HBB), which cycles into nitrogen the
carbon dredged up, if any (Iben \& Renzini 1983, Wood et al. 1983).
This interpretation has been widely confirmed by the discovery that
{\it almost all} the luminous oxygen rich AGBs in the MCs are lithium
rich, that is they show at the surface a lithium abundance $\log (\epsilon
(^7Li)) > 2$ (where $\log (\epsilon (^7Li)) = \log (^7Li/H) +12$) (Smith \&
Lambert 1989, 1990; Plez et al. 1993; Smith et al. 1995; Abia et al. 1991).
Production of lithium is possible via the so-called Cameron \& Fowler
(1971) mechanism, if the temperature at the bottom of the convective
envelope is \Tbce$\simgt 4 \times 10^7$K and non instantaneous mixing is
accounted for. Modelling of HBB started early with envelope models including
non instantaneous mixing coupled to the nuclear evolution (Sackmann et al.
1974) and the lithium production is well reproduced in the recent full models
(Sackmann \& Boothroyd 1992, Mazzitelli et al. 1999 --hereinafter MDV99--,
Bl\"ocker et al. 2000). Although the details of lithium production depend on
the input physics, mainly on the modelling of convection (MDV99), the
luminosity range at which lithium rich AGBs should appear is not sensibly
dependent on the details, and agrees well with the range observed in the
MCs, namely $-6 \simgt M_{\rm bol} \simgt -7$.
Let us recall that modelization of the lithium production in AGB is
necessary to understand the galactic chemical evolution of
lithium from the population II values $\log(\epsilon(^7Li)) \sim 2$ (Spite \&
Spite 1993; Bonifacio \& Molaro 1997) to the present solar system abundance
of $\log(\epsilon(^7Li)) \sim 3.3$ (see Romano et al. 1999 for a recent
reevaluation of the problem).

The scarciness of luminous AGBs, on the other hand, must be attributed to the
onset of strong mass loss which terminates the ``visible" evolution and leads
to a phase in which the stars are heavily obscured by a circumstellar
envelope (CSE) and eventually evolve to the white dwarf stage.
This occurs when matter of the outermost layers is found at a distance from
the star where temperature and density allow for dust formation, and
collisional coupling of the grains with the gas drives a very efficient
stellar wind (Habing et al. 1994, Ivezic \& Elitzur 1995). The stars will
then traverse a phase during which they are surrounded by a thick CSE.

A major problem in building up realistic upper AGB models is then the
modelization of mass loss, a necessary ingredient from several independent
points of view.
For the nucleosynthesis and galactic chemical evolution, the yields
from massive AGBs (in particular the lithium yields, as we will see in the
present calculations) are influenced by mass loss during the HBB phase, not
only because different amounts of processed envelopes are shed to the
interstellar medium at a given time (reflecting the stage of nucleosynthesis
which has been reached), but also because the stellar structure, and thus 
$T_{\rm bce}$  and the nucleosynthesis itself,
depends on the rate of mass loss: therefore selfconsistent models must be
explored and they have not yet been developped.

On the other hand, while at first the searches for AGB stars in the
Galaxy and the MCs had been limited to optically bright stars, the
surveys in the infrared, starting from IRAS databases or from near IR
observations, are making the cornerstones for the understanding of
the phase during which, by heavy mass loss, the objects become enshrouded in
dust, making them practically invisible at the optical wavelengths and
accessible only in the infrared (e.g. Habing 1996).
Based on these surveys (see, e.g., Zijlstra et al. 1996, van Loon et al.
1997), follow up observations have given information on the
pulsation periods; mass loss rates (\Mdot) and expansion velocity of the
envelope have been derived via the OH/IR associated masers (e.g. Wood et al.
1992); ISO spectroscopy and/or photometry has allowed to model the
bolometric luminosity and \Mdot\ (e.g. van Loon et al. 1999a).
The obscured AGBs observations will be a powerful test for the stellar
models expected to have extended CSE. 
\footnote{Actually, the carbon stars at $M_{bol} < -6$ found among
{\it obscured} AGBs in the MCs (van Loon et al. 1999b) do not change
the interpretation of the above scenario: this sample can represent 
the latest phases of AGB evolution (Frost et al. 1998), 
during which the strong mass loss cools the convective envelope 
and the dredged-up carbon is no longer cycled.}

In MDV99 we presented results from 
detailed computations focused on lithium production in massive AGB stars,
with the aim of studying the influence of convection modelling and other
physical inputs on the surface lithium abundance. Although the models were
run for population I composition stars, 
a first comparison with the LMC and SMC lithium rich AGBs
was attempted (Fig.16 in MDV99). However, a full and more detailed comparison
with the MCs requires models computed with the appropriate metallicity.
Further, in MDV99 we did not touch the problem of visibility and of a
possible calibration of mass loss.

In this paper we present stellar models starting from the pre--main sequence
and evolved to the AGB phase, with different prescriptions for \Mdot, with
the aim to describe the lithium rich AGBs observed in different luminosity
bins in the MCs bright sources. A further constraint can be put based on the
s--process enrichment observed in these same stars.

We compare the paths of evolutions having \Mdot\ consistent with these
observations with the \Mdot\ versus pulsation period
observations of obscured stars in the LMC (Wood et al. 1992) and with the
\Mdot\ versus \Mbol\ derived from ISO spectrophotometry (van Loon et al 1999a).

We finally show how the models vary with the main physical inputs.
In particular we find that, within the framework of our convection model,
the lithium vs. luminosity trend is not influenced by the overshooting
distance, but this latter is relevant to determine the range of masses
involved in lithium production, which our computations show to be 
extablished within $0.5M_{\odot}$. The minimum mass achieving HBB leading
to large lithium abundances is slightly dependent on the mass loss rate
adopted, lowest $\dot M$ models having more chances of achieving high
temperatures at the base of the external envelope during their AGB evolution.  

These new tracks represent a first numerical attempt to quantify the
mass loss in the massive AGB evolution. The mass loss parametrization
has a number of implications on the problems of population synthesis
and galactic chemical evolution, whose modeling remains very qualitative if
it is not based on such full computations.
In particular, the calibration we obtain implies that the lithium production
from the massive AGB stars is not relevant for the lithium galactic chemical
evolution.

\section {Macro Physics input}

The numerical structure of the ATON2.0 code, as well as a complete
description of the physical inputs, can be found in Ventura et al. (1998).
The description of the modeling of lithium production is in MDV99. 
We remind here the main formulations and describe the new inputs introduced
especially for this work.

\subsection {Diffusive scheme and overshooting}

Mixing of chemicals and nuclear burning are self--consistently coupled
by solving for each of the elements included in the nuclear network the
diffusive equation:

$$
{dX_i\over dt} = \left( {\partial X_i \over \partial t } \right )_{\rm nucl} +
{\partial \over \partial m} \left [ (4\pi r^2 \rho)^2 D {\partial X_i \over
\partial m} \right ]
$$

\noindent
being $X_i$ the chemical abundance of the i-th element and 
D the diffusive coefficient.

In the context of diffusive mixing also the formulation of overshooting
assumes a meaning more physically sound: we do not simply assume an
extra-mixing down to a fixed distance away from the formal convective border
(extablished according to the Schwarzschild criterium), but we allow an
exponential decay of velocity outside convective regions of the form

$$
v=v_b exp \pm \left ( {1\over \zeta f_{\rm thick}} ln {P\over P_b} \right )
$$

The exponential decay is consistent (at least on a qualitative
ground) with approximate solutions of the Navier -- Stokes equations
(Xiong 1985, Grossman 1996), and with the results of 
numerical simulations (Freytag et al.1996).
In the above expression $v_b$ and $P_b$ are the values of convective 
velocity and pressure at the convective border, $f_{\rm thick}$ is the width
of the whole convective region in fractions of $H_p$, and $\zeta$ is a free
parameter giving the e--folding distance of the exponential decay,
which we already tuned by comparing with the width of the main sequence of 
some open clusters: a conservative estimate indicates that a value 
$\zeta=0.02$ is required (Ventura et al. 1998). A value of $\zeta=0.03$ may
be required from consideration of other evolutionary phases (Ventura et al.
2000, in preparation).

\subsection{Symmetric overshooting}
The necessity of the existence of some overshooting away from convective
borders was strengtened by the difficulty of fitting the width of the main
sequences of popI open clusters without allowing any extra mixing from the
Schwarzschild border (Maeder \& Meynet 1989, Stothers \& Chin 1991): for
these hystorical reasons the term ``overshooting'' has been associated to
extra mixing from the convective core during the phases of central burning
(Meynet et al. 1993), while few models exist which take into account also
overshooting from the base of the convective envelope (Alongi et al. 1991).
For AGBs, the luminosity evolution of models 
with low core masses and overshooting below
the border of the outermost convective zone, and the possible implications
for the efficiency of the third dredge-up and the core mass - luminosity
relationship is explored by Herwig et al. (1997). 
The influence of symmetric overshooting on the evolution of the most massive
AGBs close to the limit of carbon ignition is discussed in MDV99. Further,
Ventura et al. (1999) show the possibility of reproducing the evolution
of lithium rich C-stars with a small amount of overshooting  ``from below''
the base of the external envelope. Here we do not address the issue of
the extension of such extra-mixing region, nor the changes of the surface
abundances of the elements other than lithium introduced by such extra-mixing:
we simply focus our attention on the influence of overshooting  ``from below'' 
on the lithium vs. luminosity trend in Sect.5.1

It is therefore clear that at present we cannot give a full description of
the s-process enhancement, which is the outcome of dredge-up
mixing material from the He-burning shell to the surface following each TP.
Yet, it appears evident that this enhancement can be detected only after
some TPs have taken place, and this will be a strong constrain to choose
the appropriate mass loss rate, if we consider that the luminous Li-rich  
AGBs in the MCs are all s-process enhanced.

In the following, we refer
to models without overshooting below the base 
of the envelope as ``standard''.

\subsection{Convection}

Our code can use two local models for the
evaluation of the convective flux: the mixing lenght theory (MLT) and
the Full Spectrum of Turbulence (FST) model (Canuto \& Mazzitelli 1991,1992). 
We remember that the FST model is more physically sound,
since the whole spectrum of eddies is taken into account to obtain the
convective flux, and the fluxes are consistent with experimental values
(Lesieur 1987). In addition, MDV99 showed that for the issue of
lithium production the FST gives results independent of the tuning of the
model, at variance with the MLT case, where a much larger
freedom of results can be obtained by tuning the mixing lenght parameter
(Sackmann \& Boothroyd 1992). In this work we use the FST model in its recent
version, which employs the fluxes from Canuto et al. (1996).

\subsection{Pulsation periods}
To be able to compare our theoretical tracks with the stars of
the above surveys we computed periods of our models by assuming
that variable AGB stars are pulsating in the fondamental mode,
according to Vassiliadis \& Wood (1993):
$$
\log(P)=-2.07+1.94 \log(R)-0.9 \log(M)
$$
where the period P is given in days, and the stellar radius R and mass M are
expressed in solar units.

\subsection{Mass loss}
\label{mloss}

As we have shortly seen in the introduction, mass loss has to be taken into
account in AGB computations, since AGBs may lose mass at such huge rates
($\dot M > 10^{-5} M_{\odot}$ yr$^{-1}$) (van Loon et al. 1999a) that the
general path of their evolution can be substantially altered. As no ``first
principle" approach to mass loss exists, the compromise is to try to adopt a
description based on a sound physical approach, although it needs
calibration. Our choice for the present study relies on Bl\"ocker's (1995)
formulation, which is based on Bowen's (1988) detailed numerical hydrodynamic
and thermodynamic calculations carried out for the dynamic atmospheres of
models for long period variables evolving along the AGB both in 
solar and smaller than solar metallicities (Willson et al. 1996). 
Computation of
extensive grids of models showed the extreme sensitivity of the mass loss
rate to the stellar parameters: ${\dot M}$ becomes strongly dependent on
luminosity during the AGB evolution. Bl\"ocker
(1995) starts from the usual Reimer's formulation. This is completely
inadequate to describe the fast increase of the mass loss rate during AGB,
(e.g. Habing 1996), so he introduces a further dependence on a power of the
luminosity. The complete expression is:

$$
\dot M = 4.83 \cdot 10^{-9} M^{-2.1} L^{2.7} \dot M_R
$$

\noindent
where $\dot M_R$ is the canonical Reimers rate expressed by

$$
\dot M_R=4 \cdot 10^{-13} \eta_R {LT\over M}
$$

\noindent
Tuning of the parameter $\eta_R$ is one of the goals of the present work.

The dependence of the mass loss rate on the stellar parameters is far from
being settled in Bl\"ocker formulation: it is useful to compare our main
results with other suggestions for \Mdot\ recently appeared in the
literature. Salasnich et al. (1999) provide a prescription for $\dot M$ which
has as a basis an empirical relationship between \Mdot\
and pulsation period (Feast 1991), including also the effects of a
systematic variation of the dust to gas ratio at increasing luminosities:

$$  
\log(\dot M)=2.1 \cdot \log(L/L_{\odot})-14.5
$$

The above relation depends on the luminosity only, and, if reliable,
it would make less critical the uncertainties connected with, e.g., the
convective model adopted, from which the \Teff\ of the star depends strongly
(D'Antona \& Mazzitelli 1996, MDV99)

A strong dependence on the effective temperature ($\sim T_{\rm eff}^{-8}$)
has been suggested (Schroeder at al. 1998) to trigger a large increase of
\Mdot\ at low temperatures with respect to the Bl\"ocker's recipe. Although
this formulation has been developped for Carbon stars, we will test it to
show how the strong dependence on \Teff\ counterbalances the effect of the
decreasing luminosity on \Mdot\ in the latest stages of evolution .

Remember however that these formulas are still to be considered strictly a
parametric exploratory approach.

 \begin{table*}
  \caption[]{Values of the main physical quantities during the AGB
	phase for all our models computed.}
	\medskip
   \begin{tabular*}{19cm}{cccccccccccc}   \hline
	\medskip
	$M\over M_{\odot}$ & $M_{\rm core}^{\rm 1st TP}\over M_{\odot}$ &
	$\log({L\over L_{\odot}})^a$ & $M_{\rm core}^a\over M_{\odot}$ &
	$T_{\rm max}^b$ & $\log({L\over L_{\odot}})^b_{\rm max}$ & 
	$\Delta t (^7Li)^c$ & $\Delta t (^7Li)^d$ & 
	$\log(\epsilon(^7Li))_{\rm max}$ & ${\rm Li-yield^e}$ & \\
	   \hline
	  \medskip
	  & & & & & $\eta_R=0.005$ & & & & \\
	  \hline
	  \medskip
	 3.0 & 0.620 & 4.30 & 0.779 & $6.3\cdot 10^7$ &
	       4.34  & $2.0\cdot 10^5$ & $2.0\cdot 10^5$ & 3.1 & 
	       $2.53\cdot 10^{-9}$ \\
	 3.3 & 0.696 & 4.30 & 0.782 & $6.3\cdot 10^7$ &
	       4.38  & $2.0\cdot 10^5$ & $2.0\cdot 10^5$ & 3.2 & 
	       $2.17\cdot 10^{-9}$ \\
	 3.5 & 0.755 & 4.30 & 0.794 & $6.5\cdot 10^7$ &
	       4.38  & $1.7\cdot 10^5$ & $1.7\cdot 10^5$ & 3.4 & 
	       $2.78\cdot 10^{-9}$ \\
	 4.0 & 0.874 & 4.35 & 0.881 & $7.0\cdot 10^7$ &
	       4.49  & $6.5\cdot 10^4$ & $6.5\cdot 10^4$ & 3.8 & 
	       $1.42\cdot 10^{-10}$ \\
	 4.5 & 0.917 & 4.38 & 0.918 & $7.0\cdot 10^7$ &
	       4.56  & $4.5\cdot 10^4$ & $3.9\cdot 10^4$ & 3.9 & 
	       $3.44\cdot 10^{-10}$ \\
	 5.0 & 0.959 & 4.40 & 0.959 & $7.4\cdot 10^7$ &
	       4.67  & $3.3\cdot 10^4$ & $1.7\cdot 10^4$ & 4.0 & 
	       $5.36\cdot 10^{-10}$ \\
	 5.5 & 0.998 & 4.43 & 0.997 & $7.8\cdot 10^7$ &
	       4.78  & $2.3\cdot 10^4$ & $1.0\cdot 10^4$ & 4.0 & 
	       $8.63\cdot 10^{-10}$ \\
	 6.0 & 1.050 & 4.50 & 1.030 & $8.0\cdot 10^7$ &
	       4.68  & $1.4\cdot 10^4$ & $2.0\cdot 10^3$ & 4.2 & 
	       $2.10\cdot 10^{-9}$ \\
	   \hline
	  \medskip
	  & & & & & $\eta_R=0.01$  & & & & \\
	  \hline
	  \medskip
	 3.3 & 0.705 & 4.30 & 0.782 & $6.3\cdot 10^7$ &
	       4.36  & $1.9\cdot 10^5$ & $1.9\cdot 10^5$ & 3.3 & 
	       $4.03\cdot 10^{-10}$ \\
	 3.5 & 0.750 & 4.30 & 0.793 & $6.3\cdot 10^7$ &
	       4.37  & $1.7\cdot 10^5$ & $1.7\cdot 10^5$ & 3.4 & 
	       $4.03\cdot 10^{-10}$ \\
	 4.0 & 0.876 & 4.35 & 0.878 & $7.0\cdot 10^7$ &
	       4.48  & $6.3\cdot 10^4$ & $5.5\cdot 10^4$ & 3.8 & 
	       $6.08\cdot 10^{-10}$ \\
	 4.5 & 0.918 & 4.39 & 0.918 & $7.0\cdot 10^7$ &
	       4.56  & $4.4\cdot 10^4$ & $2.6\cdot 10^4$ & 3.9 & 
	       $7.29\cdot 10^{-10}$ \\
	 5.0 & 0.960 & 4.40 & 0.960 & $7.4\cdot 10^7$ &
	       4.61  & $2.8\cdot 10^4$ & $6.7\cdot 10^3$ & 4.0 & 
	       $1.29\cdot 10^{-9}$ \\
	 5.5 & 1.002 & 4.42 & 1.001 & $7.9\cdot 10^7$ &
	       4.69  & $2.2\cdot 10^4$ & $5.7\cdot 10^3$ & 4.0 & 
	       $1.89\cdot 10^{-9}$\\
	 6.0 & 1.050 & 4.50 & 1.050 & $8.0\cdot 10^7$ &
	       4.75  & $1.4\cdot 10^4$ & $1.0\cdot 10^3$ & 4.2 & 
	       $4.10\cdot 10^{-9}$ \\
	   \hline
	  \medskip
	  & & & & & $\eta_R=0.05$  & & & & \\
	  \hline
	  \medskip
	 3.3 & 0.702 &  -   &   -   &$3.2\cdot 10^7$ &
	       4.27 &      -           &     -           & 0.0 & 
	       $1.76\cdot 10^{-11}$ \\
	 3.5 & 0.750 & 4.30 & 0.795 & $6.3\cdot 10^7$ &
	       4.36  & $8.0\cdot 10^4$ & $5.7\cdot 10^4$ & 3.2 & 
	       $4.26\cdot 10^{-10}$  \\
	 4.0 & 0.870 & 4.34 & 0.876 & $7.0\cdot 10^7$ &
	       4.47  & $6.2\cdot 10^4$ & $1.5\cdot 10^4$ & 3.7 & 
	       $2.73\cdot 10^{-9}$\\
	 4.5 & 0.915 & 4.40 & 0.916 & $7.0\cdot 10^7$ &
	       4.54  & $4.0\cdot 10^4$ & $7.0\cdot 10^3$ & 3.9 & 
	       $3.51\cdot 10^{-9}$ \\
	 5.0 & 0.960 & 4.60 & 0.960 & $7.0\cdot 10^7$ &
	       4.67  & $2.7\cdot 10^4$ & $4.0\cdot 10^3$ & 4.0 & 
	       $5.78\cdot 10^{-9}$\\
	 5.5 & 1.000 & 4.50 & 0.996 & $7.9\cdot 10^7$ &
	       4.75  & $1.9\cdot 10^4$ & $3.0\cdot 10^3$ & 4.0 & 
	       $7.33\cdot 10^{-9}$\\
	 6.0 & 1.050 & 4.50 & 1.052 & $8.0\cdot 10^7$ &
	       4.75  & $1.3\cdot 10^4$ & $5.0\cdot 10^2$ & 4.2 & 
	       $1.49\cdot 10^{-8}$ \\
	  \hline
	 \medskip
	 & & &  & & $\eta_R=0.1$  & & & & \\
	 \hline
	 \medskip
	 3.5 & 0.751 &  -   &  -    & $3.5\cdot 10^7$ &
	       4.28 &     -         &      -        & 0.2 & 
	       $1.43\cdot 10^{-11}$ \\
	 3.7 & 0.804 & 4.30 & 0.820 & $6.0\cdot 10^7$ &
	       4.36 & $6.0\cdot 10^4$ &     -       & 3.1 & 
	       $8.08\cdot 10^{-10}$ \\
	 4.0 & 0.875 & 4.35 & 0.879 &  $7.0\cdot 10^7$ &
	       4.46 & $5.2\cdot 10^4$ &     -       & 3.8 &  
	       $4.94\cdot 10^{-9}$ \\
	 4.5 & 0.917 & 4.39 & 0.918 &  $7.0\cdot 10^7$ &
	       4.54 & $3.4\cdot 10^4$ & $3.0\cdot 10^3$ & 3.9 &
	       $6.49\cdot 10^{-9}$ \\
	 5.0 & 0.961 & 4.43 & 0.961 &  $7.0\cdot 10^7$ &
	       4.60 & $2.5\cdot 10^4$ & $2.0\cdot 10^3$ & 4.1 &  
	       $9.61\cdot 10^{-9}$ \\
	 5.5 & 1.030 & 4.44 & 1.038 &  $7.9\cdot 10^7$ &
	       4.72 & $1.3\cdot 10^4$ &     -        & 4.1 &  
	       $1.69\cdot 10^{-8}$ \\
	 6.0 & 1.052 & 4.50 & 1.050 &  $7.9\cdot 10^7$ &
	       4.75 & $1.0\cdot 10^4$ &     -        & 4.2 &  
	       $2.31\cdot 10^{-8}$ \\
	    \hline
   \end{tabular*}
\begin{list}{}{}
\item[$^{\mathrm{a}}$] Values of luminosities
     and core masses refer to the beginning of the ``super--rich''
     phase, when $\log(\epsilon(^7Li))\geq 2$.
\item[$^{\mathrm{b}}$] These values refer to the time when lithium
     abundance is at its maximum value.
\item[$^{\mathrm{c}}$] Total duration (in years) of the phase when 
     $log(\epsilon(^7 Li)) \geq 2$.
\item[$^{\mathrm{d}}$] Total duration (in years) of the phase when 
     $log(\epsilon(^7 Li)) \geq 2$ and $\tau_F\leq 0.3$.
\item[$^{\mathrm{e}}$] The lithium yield is defined as the ratio between
the total amount (in mass) of lithium produced within the star
and ejected in to the interstellar medium and the initial mass of the
star.
\end{list}
 \end{table*}

\subsection{Model observability}
The models we build up provide as observables the bolometric luminosity,
the photospheric \Teff\ (obtained through a grey atmosphere integration) and
\Mdot. If we wish to compare the results to selected samples of stars, we
need to compute bolometric corrections and colors and, when \Mdot\ becomes
important, we have to worry about the presence of an optically thick CSE,
and should compute how the photospheric quantities are modified by the
envelope. Although this kind of approach is in preparation (Groenewegen \&
Ventura 2000 in preparation), 
for the present work we mainly need to understand whether the
models we build up correspond to stars which emit a good fraction of light
into near IR wavelengths (so that they could be easily discovered in the K
band surveys) and whether the optical red part of the spectrum --including
the lithium line-- is observable. When the mass loss becomes too large, the
models represent more obscured phases, and should be compared with samples of
objects whose near IR colors become very red, and whose optical spectroscopy
becomes difficult or impossible with present day instrumentation (e.g. Garcia
Lario et al. 1999). In particular, our main test will be made by comparing
with the Smith et al. (1995) stars, which are luminous, non obscured AGBs,
for which CSE absorption is probably negligible, as it is indicated by
their near IR colors.

We compute for our models the flux-weighted optical depth \tauf\ (e.g. Ivezic
\& Elitzur 1995), defined as the ratio between the mass loss rate and the
classic rate based on the single-scattering approximation, given by the
condition that the momentum flow of the gas ($\dot M v_{\rm exp}$) 
equals that of the photons ($L/c$). 
The classic value is thus $\dot M_{classic}=L/(v_{exp}
\cdot c)$, whereas the flux-weighted optical depth \tauf\ is given by

$$
\tau_F=\dot M \cdot v_{\rm exp} \cdot c/ L
$$

Habing et al. (1994) have stressed the possibility of having stars
with momentum flow much larger than the gas momentum. To find out the value
of $\tau_F$ we rely on the results of previous computations, which
indicate a dependency of $v_{\rm exp}$ on luminosity of the form
$v_{\rm exp} \sim L^{0.25}$ (Jura 1984; Habing et al 1994). In order to
have a calibration adequate to the LMC metallicity the constant of
proportionality has been fixed by demanding a star with $L=30000L_{\odot}$
to have $v_{\rm exp} = 10$ Km $s^{-1}$ (van Loon et al. 1999a). We therefore 
approximate \tauf\ by

$$
\tau_F=3.64\cdot 10^7 \dot M (L/L_{\odot})^{-0.75}
$$

This is a first order approximation to find out likely values
for a physically sound parameter. We have to caution that the LMC expansion
velocities are very uncertain (only a few determinations based on noisy data)
and the derivation of these velocities also depend on the mass loss rate
(Steffen et al. 1998).

We computed $\tau_F$ along the evolution of our
models. From test computation of the emerging fluxes from the CSE for our
models (Groenewegen \& Ventura 2000 in preparation), 
we can make a rough division between
models: if \tauf$\simlt 0.3$, we can consider the model not much affected by
the CSE and include it in the comparison with the lithium data from the
Smith et al. (1995) samples, while for
\tauf$\simgt 0.3$\ the CSE increasingly dominates and the models are to be
compared with the samples of obscured stars.

\section{Model results}
The main aim of this work is to construct a simple population
synthesis adequate to describe the Magellanic Clouds upper AGB stars, in order
to derive information on the mass loss formulation, which is a physical input
very poorly constrained by theory. Consequently, following MDV99 which
provides a general description of the new physical inputs adopted in our
models and of the detailed results, we now build up models adequate for the
chemical composition of the LMC, namely Z=0.01.

\begin{figure}
\caption[]{Evolution in terms of lithium production of some of our
intermediate mass models: upper panels refer to computations performed
with $\eta_R=0.01$, lower panels correspond to the case $\eta_R=0.05$.}
\label{fig_all}
\end{figure}

The models cover the mass range $3M_{\odot} \leq M \leq 6.5M_{\odot}$ with
mass steps of $0.5M_{\odot}$. We built four sets of models corresponding to
the values for the parameter $\eta_R$ in the mass loss formula: $\eta_R=$
0.005, 0.01, 0.05 and 0.1. The overshooting parameter was set to
$\zeta=0.02$. In all cases only overshooting from the external border of
convective cores is considered; the influence of symmetric overshooting will
be discussed in Sect. \ref{symm}. Table 1 lists the computed models and
some of the interesting physical quantities. Notice first that the
$6.1M_{\odot}$ models ignites carbon at the centre of the star in a
semi-degenerate regime, thus jumping the phase of thermal pulses.
The Table confirms the results obtained by MDV99, and is also
consistent with other authors recent main results (Sackmann \& Boothroyd 1992,
Bl\"ocker et al. 2000, Forestini \& Charbonnel 1997), once the
differences in the approach to convection, mass loss and the different
chemistry are taken into
account. In the following we describe the feature of models as a
function of the mass loss rate, the main input which we are trying to
constrain.

\subsection{Mass and Luminosity at which HBB is achieved}
The minimum mass evolving as a Li-rich AGB goes from $3M_{\odot}$
($\eta_R=0.005$) to $3.7M_{\odot}$ ($\eta_R=0.1$). A larger mass loss rate in
fact triggers an earlier cooling of the outer layers of the star before the
ignition of HBB. The minimum core mass required for lithium production
ranges from $\sim 0.78$ to $\sim 0.82M_{\odot}$, but the corresponding
luminosity is always $\log(L/L_{\odot}) \simeq 4.3$; we therefore expect to
find lithium rich sources for \Mbol $\leq -6$, in excellent agreement with
the results of the survey by Smith et al. (1995) and independently from
the mass loss rate.

\subsection{Lithium yields}
Fig.\ref{fig_all} shows the evolution of surface lithium abundance for some
models computed with $\eta_R=0.05$ and $0.01$. Note that the $6M_{\odot}$
model achieves temperatures sufficient to ignite the Cameron--Fowler
mechanism well before the beginning of thermal pulses, so that when the first
pulse starts the lithium abundance is already in the declining branch. A
similar behaviour is found in the $M=5.5M_{\odot}$ case (not shown).

The largest amount of surface lithium which these stars can produce is a
slightly increasing function of initial mass. The phase during which the star
shows up lithium rich is obviously shorter for the models of larger $\eta_R$,
since the larger mass loss rate causes a decline of luminosity and a cooling
of the whole external envelope, thus turning off the hot bottom burning.

The last column in Table 1 shows the lithium yield. The
masses close  to the limit of carbon ignition provide the major contribution,
due to the fact that the $5.5,6M_{\odot}$ models begin lithium production
while the luminosity is still rising, so that the maximum lithium abundance
is achieved in conjunction with the largest value of luminosity (and hence of
mass loss). For the lowest masses, we note from Table 1 that the
yield of the $3.5M_{\odot}$ models of $\eta_R=0.005$ are not due to lithium
production, but to the survival of lithium from the previous evolutionary
phases. As the present abundance of lithium in the ISM is $\log X_{\rm Li}
\simeq -8$, starting from the population II abundance of $\simeq -9$, only
the models with $\eta_R \simgt 0.05$ can significantly contribute to the
galactic enrichment of the ISM, should they be consistent with the other
evolutionary constraints.

\subsection{Mass loss and the duration of the optically bright phase}
\begin{figure}
\caption[]{Evolution of some intermediate mass models computed with
$\eta_R=0.005, 0.01, 0.05$\ in the \Mbol $- \log(age)$ plane. 
Times were normalized at the beginning of the AGB phase.
Triangles along the tracks indicate the location of the first thermal pulse,
full circles point the stage at which the optical depth $\tau_F$ exceeds
0.3. Note that according to this criterium the
$3.5M_{\odot}$ models computed with the lower values of $\eta_R$ never 
become invisible in the optical.}
\label{fig_tautime}
\end{figure}

\begin{figure}
\caption[]{Evolution of some intermediate mass models computed with
$\eta_R=0.005, 0.01, 0.05, 0.1$\ in the \Mbol $- \log \tau_F$ plane. For
$\eta_R=0.005, 0.01$ we report tracks corresponding to initial masses $M=3.5,
4, 4.5 M_{\odot}$, while for $\eta_R=0.05, 0.1$ we show $M=4, 4.5,
5M_{\odot}$. The heavy-dashed track gives 
the evolution of a $4.5M_{\odot}$ model
computed by adopting the Schroeder et al. (1998) mass loss rate from the
point when the luminosity was at the top. Continuous horizontal line
indicates the threshold value $\tau_F=0.3$ above which we assume that the
star can not be detected in the optical.}
\label{fig_tauf}
\end{figure}

Fig.\ref{fig_tautime} shows the time spent in the AGB phase for different
masses and $\dot M$. The visible phase for each mass ends in correspondence
of the full point along the track. 
The evolution of the flux-weighted optical depth (\tauf) along
our sequences with different $\eta_R$\ is shown in Fig.\ref{fig_tauf}. The
important evolutionary region is at $-6 \simgt$ \Mbol $\simgt -7$\ where the
bright lithium rich AGBs are located.
The mass to be attributed to these stars depends on the mass loss rate:
If $\eta_R=0.01$\ or 0.005, the sequences from 3.5 to 4.5\Msun\  evolve
almost completely below $\tau_F=0.3$,
so that these models should never be particularly obscured, and we can
expect that these are the masses which populate the AGB. On the contrary,
if $\eta_R$\ is as large as 0.05 or 0.1, the sequences below 5\Msun
 apparently are already obscured when they populate 
the most luminous bin $-6.5 \simgt$ \Mbol $\simgt -7$.
For all sequences the evolution proceeds from low \tauf\ to larger values, at
which the CSE dominates. In the end, \tauf\ may decrease again due to the
decrease in the total stellar luminosity and consequent decrease in \Mdot. In
the cases $\eta_R=0.005$ and 0.01 the models evolve considerably in
luminosity at low \tauf, that is when they have no appreciable CSE, while for
the larger $\eta_R$\ cases the part of the sequence not affected by the CSE
is only a small fraction. In other words, the
larger is \Mdot, the shorter is the TP phase during which the
model represents a luminous AGB star corresponding to the Smith et al. (1995)
sample. In particular, the 6\Msun\ sequence of $\eta_R=0.05$ and 0.1 would
evolve into a dust embedded phase even {\it before} they start the thermal
pulse phase. Can we find a way of discriminating, at least qualitatively,
between the four rates? We will try to do this using the luminosity and
number distribution of lithium rich AGBs.

\begin{figure}
\caption[]{Comparison between the variations with core mass
of some physical and chemical quantities of two
$4M_{\odot}$ models calculated by assuming two different mass loss
rates, corresponding to the Bl\"ocker's recipe with $\eta_R=0.01$
and $\eta_R=0.005$. The thin dashed line in the bottom panel
indicates the value of $\tau_F$ we selected to separate stars still
observable in the optical from those heavily obscured. We see that in 
the $\eta_R=0.005$ case the optical depth 
keeps low up to phases when $\log(\epsilon(^7Li))$ has dropped below 0
(middle panel).}
\label{fig_mloss1}
\end{figure}

\subsection{Mass loss and Lithium evolution}
Fig.\ref{fig_mloss1} shows the effect of mass loss on the luminosity
evolution of two $4M_{\odot}$ models computed with
$\eta_R=0.01, 0.005$. The variation with core mass of luminosity, lithium
and optical depth $\tau_F$ are reported. We see that the evolution in
luminosity is slightly different: in the $\eta_R=0.01$
case $\log(L/L_{\odot})$ attains a maximum value of 4.67 when the core mass
is $M_{\rm core} \sim .904M_{\odot}$. In the $\eta_R=0.005$ model the luminosity
attains its maximum of $\log(L/L_{\odot}) =4.71$ when
$M_{\rm core}=0.908M_{\odot}$. At this point the difference
between the bolometric magnitudes of the two models is 0.13 mag.

The bottom panel shows that for a fixed value of the core mass (hence,
of luminosity) the optical depth in the $\eta_R=0.005$ case is approximately
half the value of the $\eta_R=0.01$ case, making the star visible for a
longer time.
The main difference between the two models is therefore that {\it in the
$\eta_R=0.005$ case the star remains visible in the optical until phases when
the lithium abundance has already dropped to low values, due both to the very
large $T_{\rm bce}$'s and to the exhaustion of $^3He$ in the external envelope}.
Observationally we would expect in the latter case to detect several large
luminosity sources with negligible amounts of lithium.

\section{Numeric simulations of the optically bright AGB phase}

To have more quantitative informations, we compute simple population
synthesis for a sample of stars which go through the HBB phase described. We
assume an IMF with exponent $-2.3$ for the mass distribution and consider
that the range of evolving masses is limited in between 3.0 and 6\Msun.
The evolutionary tracks are considered only until the models have
\tauf$\simlt 0.3$. We extract randomly the value of the initial mass and then
allocate it at an age (an thus a luminosity) chosen randomly again in the
time interval between the beginning of the AGB phase and the $\tau_F=0.3$
time (this is equivalent to assume a constant birthrate between now and the
time at which the lowest masses considered - $3M_{\odot}$ - evolve, i.e.
$4\cdot 10^8$ yr).

We show in Fig. \ref{histo} the result of our simulations for the ratio of
lithium rich AGBs versus their total number in intervals of 0.5 mag in \Mbol,
compared with the Smith at al. (1995) data (from their Fig.9). The comparison
must be limited to the bins of $-6 \simgt M_{\rm bol} \simgt -7$, which
correspond to the models which manifacture lithium (in fact
the bins at \Mbol $\geq -6$ in our models correspond to the early AGB phases,
during which the temperature at the base of the envelope was not large enough
to destroy the surface lithium remnant of the previous evolution, while
the stars at \Mbol $\geq -6$\ in Smith et al. (1995) show strong evidences
of s-processes enrichment, and are thus likely a result of AGB evolution of
sources with initial masses below $M_{ZAMS} \sim 3M_{\odot}$, which are not
considered here, so the good agreement with our models at these magnitudes is
fortuitous).

\begin{figure}
\caption[]{Frequecy histogram of lithium rich AGBs as a function of \Mbol,
for three sets of models with different mass loss rates. Dots indicate the
frequencies related to the survey of the most luminous AGBs in the MCs by
Smith et al. (1995). In the smallest \Mdot\ case ($\eta_R=0.005$) we
would expect the majority of AGB sources with $-7 \leq$ \Mbol $\leq
-6.5$ to have negligible surface lithium, leading to a strong discrepancy
with the observational evidence.}
\label{histo}
\end{figure}

\begin{figure}
\caption[]{Numerical simulations for the distribution of our AGB models
in the magnitude - mass plane, obtained by assuming that the sources
become invisible in the optical as soon as $\tau_F$ becomes equal to 0.3.
Results for $\eta_R=0.005, 0.01, 0.05$ are reported. Also shown are the 
lines indicating the first thermal pulse for each mass (dotted), the 
points where Li-rich abundances are achieved (dashed), and the limit of
detectability in the optical (long-dashed). The relatively low number
of stars with masses $M > 4M_{\odot}$ is due to the difference among the
various models in terms of AGB life times; for the sake of clarity 
a flat mass function has been adopted in building this figure.}
\label{simul}
\end{figure}

The main result of Fig.\ref{histo}\ is that the lowest mass loss rate
($\eta_R=0.005$) predicts that only about 20\% of AGBs in the bin -6.5 $\simgt
M_{\rm bol} \simgt -7$ should be lithium rich, while in the Smith et al.
sample almost all these stars are lithium rich. (In fact there are two
SMC stars without lithium at $M_{\rm bol} \sim -7$. This is included in the
poissonian error bar of our Fig.\ref{histo}).

The result can be understood by looking at Fig.\ref{simul}, where we report
the outcome of our simulations made for the three mass loss rates adopted
considering a flat mass function and taking into account all masses
$3M_{\odot} \leq M_{ZAMS} \leq 6M_{\odot}$. The flat IMF has no incidence
upon our main findings in terms of the fraction of Li-rich stars found in the
various bins of magnitude, and has the advantage of showing up more clearly
the results.

In Fig.\ref{simul} we also show the line
at which TPs begin (dotted), where the stars become
Li-rich (dashed line), and where $\tau_F$ reaches the value of 0.3
(long dashed).
We consider first the $\eta_R=0.005$ case (top panel of Fig.\ref{simul}).
Here the $3.5M_{\odot}$\ model reaches luminosities \Mbol $\leq -6.5$ before
strong mass loss leads to a decline in the luminosity. In these final stages
of the evolution the surface lithium is exhausted because of the lack of
$^3He$ within the envelope, so that, considering the longer life-times of the
$3.5M_{\odot}$ model, we expect that the majority of stars at $-7 \leq$ \Mbol
$\leq -6.5$ have no lithium, as in fact shown in Fig.\ref{histo}. The mass
loss rate corresponding to such $\eta_R$ seems therefore to be too low.

In the $\eta_R=0.01$ case (middle panel of Fig.\ref{simul}) the rise of 
luminosity in the $3.5M_{\odot}$ model is halted by the mass loss in the
final stages of the evolution, therefore stars with magnitudes $-7 \leq$
\Mbol $\leq -6.5$ are the descendants of masses $M\geq 4M_{\odot}$. Note
that, within this framework, stars with initial masses $M\geq 5.5M_{\odot}$
populate this region before the beginning of TPs: these latter sources,
however, would constitute less than $5\%$ of the whole population.

In the $\eta_R=0.05$ case, mass loss is so large that only masses $M\geq
5M_{\odot}$ can ever reach stages where \Mbol $\leq -6.5$. But these stars
would cross this interval of magnitudes well before the first TP, as can be
seen following the dotted line in Fig.\ref{simul}. Although modelling of the
third dredge-up is still uncertain, it is qualitatively necessary that the
stars suffer TPs to have the possibility to dredge-up s-process elements, as
already discussed in Sect.2. Therefore the stars corresponding to $M\sim 5.5
- 6M_{\odot}$ would not display any evidence of s-process elements
enrichment; this is in contrast with the observations of AGBs in the MCs,
which show enrichment of s-process elements in all the Li-rich sources. On
the basis of this discussion we may conclude that $\eta_R<0.05$ is required.

The case $\eta_R=0.01$ seems to provide the best agreement between our models
and the observations of AGBs in the MCs. This would indicate that the most
luminous Li-rich AGBs in the MCs are the descendants of stars with initial
masses $M\sim 4 - 4.5M_{\odot}$.
In MDV99 we had identified these most luminous stars with the evolution
of the 6\Msun\ models: this was due to our neglect of the ``visible" phase,
and also to the neglect of the information deriving from s--process
enhancement in the MC stars.
 
One uncertain point in the above discussion is the threshold value of
$\tau_F^{\rm max}=0.3$ at which we assume that the stars become invisible in the
optical. We tested the sensitivity of our main conclusions to the choice of
such $\tau_F^{\rm max}$.
A variation by 0.1 in $\tau_F^{\rm max}$ shifts the long dashed line in
Fig.\ref{simul} horizontally by about 0.15 mag. 
The case $\eta_R=0.005$ case can be ruled
out anyway, since the evolution of the $3.5M_{\odot}$, which  is the main
contributor to the population at $-7\leq$ \Mbol $\leq -6.5$, has $\tau_F <
0.1$. Therefore even lowering $\tau_F^{\rm max}$ by $50\%$ we still 
would expect the majority of the most luminous AGBs to be without lithium.
In the $\eta_R=0.05$ case the major difficulty is to populate
the region $-7\leq$ \Mbol $\leq -6.5$ for stars with masses $M < 5M_{\odot}$:
this problem still holds by varying $\tau_F^{\rm max}$, 
because $\dot M$ attains so large values that masses $M\sim 4 - 4.5M_{\odot}$ 
cannot reach such luminosities anyway. 
The evolution of the more massive models is in contrast with the presence
of s-enrichment; in order to have their evolution observable well after
the beginning of the TP phase, $\tau_F^{\rm max}$ should be $\geq 0.6$, value
really very large to be compatible with the ``normal'' colors  of the
Smith et al. (1995) sample.
What can be said for $\eta_R=0.01$?
In this case the agreement with the observations is due to the fact that the
evolution of the $3.5M_{\odot}$ model never reaches luminosities as large as
\Mbol $\sim -6.5$. The last bin in Fig.\ref{histo} would be mainly 
populated by stars with $M_{ZAMS} \sim 4 - 4.5M_{\odot}$, in a
phase when they are Li-rich. By adopting a smaller $\tau_F^{\rm max}$ we expect
on the average a lower abundance of s-process elements, but the
percentage of Li-rich luminous AGBs would be close to $100\%$. Our numerical
simulations show that the results are almost completely unchanged if we
`adopt a larger $\tau_F^{\rm max}$, because in this latter case the same sources
would be observable up to later stages when their luminosity exceeds 
\Mbol $\sim -7$, thus populating a region out of the limits
$-7 \leq$ \Mbol $\leq -6.5$. No such luminous AGBs 
have been observed in the quoted survey, so that a  
value of $\tau_F^{\rm max}$ largely exceeding 0.3 is probably to be excluded.

\begin{figure}
\caption{Evolution of some $\eta_R=0.005, 0.01, 0.05$ models in the $\log(P) -$
\Mbol  plane (right), where P is the period expressed in days. Full
triangles along the tracks indicate the first thermal pulse; 
open points and
crosses refer to the AGB (points) and supergiant (crosses) stars in the LMC
sample given by Wood et al. (1983); full squares correspond to the sample of
obscured AGB stars in the LMC found in Wood et al. (1992).}
\label{fig_wood}
\end{figure}

If our calibration of $\eta_R$ is valid, from Table 1
we easily recognize that these AGBs can not 
influence in any appreciable way the galactic increase of the lithium
abundance from its population II value ($\log(\epsilon(^7Li)) \sim 2.2$)
to the popI standard value of $\log(\epsilon(^7Li)) \sim 3.1$. Full
consideration of this problem will be given in a following work
(Ventura \& Romano, in preparation). 

\section{The obscured phase}
\subsection{Pulsation period evolution}
The computations so far performed allow us further comparisons with other sets
of data. The evolution of our models in the $\log(P)-$ \Mbol\ plane is shown
in Fig.\ref{fig_wood} together with the samples by Wood et al. (1983) and
Wood et al. (1992). Fig.\ref{fig_wood} shows that the evolutionary sequences
first cover the luminous long period variables region (Wood et al. 1983, open
circles) and then evolve to longer periods, where they match the location of
OH/IR stars in the Wood et al. (1992) sample, at periods above 1000d (filled
squares). The sequences with initial masses $4M_{\odot} \leq M \leq
4.5M_{\odot}$\ traverse the region $-7\leq$ \Mbol $\leq -6.5$, which correspond
to the last bin in Fig.\ref{histo}. For $\eta_R=0.01$ their \tauf\ does not
exceed 0.3 (see Fig. \ref{fig_tauf}), and so we can expect that the models
describe stars which are not particularly obscured.
The most massive AGB sequences ($M_{ZAMS} = 6M_{\odot}$) with low $\eta_R$\
reach luminosities as large as \Mbol $\sim -7.5$, at periods exceeding 1000
days, and could represent the two large luminosity, very long period stars
shown. In fact, the evolution of the $6M_{\odot}$ models not only reaches
such a luminous \Mbol , but has a large $\tau_F$,
pointing to stars with strong
CSE. Certainly, we predict that these two stars are lithium rich, 
but they are probably not observable in the red part of the visible spectrum.

\subsection{Other mass loss formulations}

Fig.\ref{fig_tauf} shows that the somewhat simple minded parameter \tauf\
is not costantly increasing during the evolution. It may well be that in
some cases the decrease of the stellar luminosity leads to such a reduction
of \Mdot\ that the objects becomes less obscured. Stop of HBB can in that
case also lead to the late formation of a carbon star (Frost et al. 1998).
The reduction in \tauf\ depends on the mass loss formulation
we have adopted. We shortly show comparison with other formulations.
The evolution of a $4.5M_{\odot}$ model has also been
computed by applying a correction of the form \Teff $^{-8}$ to 
Bl\"ocker's recipe with $\eta_R=0.05$ (heavy-dashed line in
Fig.\ref{fig_tauf}): the net result is that the effect of the decline of
luminosity is completely counterbalanced by the decrease of \Teff , so
that the optical depth of the star is still large. Thus the large mass loss
rates obtained with $\eta_R=0.1$ can be also obtained with a strong
dependence on \Teff.

\begin{figure}
\caption[]{Variation with luminosity of the mass loss rate (expressed in
$M_{\odot}/yr$ units) of some models computed with different
prescriptions for $\dot M$. In the $\eta_R=0.005, 0.01$ cases we show
the evolution of masses $3.5M_{\odot}\leq M \leq 5M_{\odot}$, for
$\eta_R=0.05, 0.1$ we report tracks for $M=4, 5, 6M_{\odot}$. 
The heavy line corresponds to a $5M_{\odot}$ model computed by adopting the
Salasnich et al. (1999) formula. Open points indicate the results concerning
LMC sources given in van Loon et al. (1999a).}
\label{fig_mloss2}
\end{figure}

We finally tested the Salasnich et al. (1999) recipe for mass loss.
Fig.\ref{fig_mloss2} shows the variation with luminosity of various
models computed with $\eta_R$ in the range $0.005 - 0.1$ and a $5M_{\odot}$
model computed with the Salasnich et al. (1999) recipe. 
We can clearly distinguish the different
slope of the latter prescription with respect to the others. 
The corresponding $\dot M$, particularly at large luminosities, 
turns out to be too low: a $3.5M_{\odot}$ evolution
would last in the non-obscured phase for $\sim 10^5$ yr at 
$M_{\rm bol} \sim -6.5$
after all lithium has been already burned (like in the $\eta_R=0.005$ case), 
in contrast with the observations.

\subsection{Comparison with other mass loss informations}
Fig.\ref{fig_mloss2} shows values of $\dot M$ by detailed computations 
by van Loon et al. (1999a), based
on IR observations of several sources in the LMC (Schwering \& Israel 1990;
Reid et al. 1990): we should remember of course that the 
observed $\dot M$'s have several uncertainties connected with the
assumptions made concerning the expansion velocities and the
dust to gas ratio, and are time averaged due to the IR photometry. 
A precise fit cannot be expected. Also, $M_{\rm bol}$ should be
treated with some care since the distance to the LMC is subject to
discussion (published range covers 0.4 mag at present) and $M_{\rm bol}$
is itself uncertain by $0.1-0.2$ mag for variable, red stars.

Values of $\eta_R=0.1$ seem to be a very upper limit: 
our $6M_{\odot}$ model computed with such $\eta_R$
attains values of $\dot M$ which in some cases exceed the largest
observed values.
The large spread of the points in Fig.\ref{fig_mloss2}
shows the difficulty of fitting the observational evidence with
a single mass loss rate, suggesting a possible spread of values of $\eta_R$.

It is important here to remember that these values for $\eta_R$ apply to
the relationships $M_{\rm core} - M_{ZAMS}$ and $M_{\rm core}$ - luminosity 
provided by {\it our own models}. Our models provide the largest core masses
in the literature (Wagenhuber \& Groenewegen 1998), and use of the FST
model of turbulent convection leads to a steeper core mass - luminosity
relationship with respect to MLT models (D'Antona \& Mazzitelli 1996):
due to the steep dependence of Bl\"ocker's recipe on luminosity, both these
effects lead to larger mass loss rates in our models during the AGB phase. 
MLT models would require larger values of $\eta_R$. 
The conclusions on the lithium yields, however, would remain valid.

\section{Overshooting}
\subsection{Symmetric overshooting}
\label{symm}

Models computed in the present work do not include any extramixing from the
base of the external envelope. Here we focus our attention on how
``overshooting from below" might change our results.
The larger extension
of the inward penetration of the convective envelope following each pulse,
in the symmetric overshooting models, brings some helium at the surface
of the star, thus delaying the ignition of the following pulse.
The interpulse phase is $\sim 3$ times longer.
The delay in the occurrence of thermal pulse causes the pulse
to be ignited at larger temperatures, so that its strength is enhanced. 

\begin{figure}
\caption[]{Evolution of surface chemical abundances of carbon
and lithium for the $4M_{\odot}$ models computed
by assuming just overshooting from the convectice core of the star
during the phases of central nuclear burning (solid track), or also
symmetric overshooting from the base of the external convective
envelope (dotted).}
\label{fig_ovsimm}
\end{figure}

Fig.\ref{fig_ovsimm} shows the variation with time of the surface abundances
of carbon and lithium during the AGB phase for both the standard and the
symmetric overshooting model of initial mass $M=4M_{\odot}$. 
Overshooting from below can no
longer be neglected if we are interested in the surface abundance of elements
like $^{12}C$, which are carried to the stellar surface from internal layers
during the third dredge--up: in particular the ``carbon star'' phase cannot
be achieved by the standard model. Although it is not yet monitored in
our models, the third dredge-up will also be necessary to bring at the
surface the s-process elements manifactured inside the star by the
$^{13}C+\alpha$ neutron source (a $^{13}C$ pocket is naturally created by our
overshooting treatment, see MDV99). The lithium rich AGBs in the MCs are
s-process enhanced, although the enhancement decreases with the luminosity,
as it would be expected if higher masses, 
suffering a smaller number of thermal pulses and a
smaller number  of dredge-up episodes, populate the high luminosity regions.
The third dredge-up is very much model dependent, and no agreement
yet has been reached among researchers on its modalities (Lattanzio 1986;
Straniero et al. 1997; Herwig et al. 1997), mainly
because overshooting can not be modelled by first principles.
On the contrary, the bottom panel of Fig.\ref{fig_ovsimm} shows
that the evolution  of the surface lithium abundance  is unchanged in
models including overshooting, so that lithium production and 
its correlation with luminosity, which we have used to
calibrate the mass loss, can be regarded as a robust result,
independent of the inclusion or not of overshooting from below.

\subsection{The influence of $\zeta$}

The results given in the previous section, particularly for that concerning
the interval of initial masses which are involved in lithium production, are
dependent at a certain extent on the amount of overshooting assumed from the
Schwarzschild border of convective cores during phases of central burning,
i.e. on the value of the parameter $\zeta$. The way the results change with
$\zeta$ is straightforward: a larger overshooting distance leads to larger
core masses at the beginning of TPs, and, for a given initial mass, the
probabilities of ignition of the Cameron--Fowler mechanism increases.
Consequently, a larger $\zeta$ would shift downwards the interval of masses
given in the previous section.

 \begin{table}
  \caption[]{Values of some physical quantities of the evolution
  of $4M_{\odot}$ models computed with different values of the
  overshooting parameter $\zeta$.}
	\medskip
   \begin{tabular*}{9cm}{cccccccc}   \hline
	\medskip
	$\zeta\over H_p$ & $M_{\rm core}^{\rm 1st TP}\over M_{\odot}$ &
	$\log({L\over L_{\odot}})^a$ & $M_{\rm core}^a\over M_{\odot}$ &
	$\log({L\over L_{\odot}})_{\rm max}$ & 
	$M_{\rm core}^b\over M_{\odot}$ & \\ 
	   \hline
	  \medskip
	 0.00 & 0.775 & 4.30 & 0.802 & $4.40$ & 0.835 \\
	 0.02 & 0.870 & 4.34 & 0.870 & $4.45$ & 0.890 \\
	 0.03 & 0.880 & 4.34 & 0.885 & $4.45$ & 0.900 \\
	   \hline
   \end{tabular*}
\begin{list}{}{}
\item[$^{\mathrm{a}}$] Values of luminosities
     and core masses at the beginning of the 
     phase when $\log(\epsilon(^7Li))\geq 2$.
\item[$^{\mathrm{b}}$] Values at which $\log(\epsilon(^7Li))$
      declines again below 2. 
\end{list}
 \end{table}

To quantify the sensitivity of the results obtained on the value of $\zeta$
assumed for the present computations we compare in Table 2
the results of three $4M_{\odot}$ evolutions computed, 
respectively, with $\zeta=0, 0.02$ and $0.03$.
We see that there is a difference of about $\sim 0.1M_{\odot}$ between
core masses at the first TP of the $\zeta=0$ and $\zeta=0.02$ cases, while
the difference between the two overshooting models corresponding to
$\zeta=0.02$ and $\zeta=0.03$ is $\sim 0.01M_{\odot}$. This 
result indicates that a variation of $50\%$ of the
overshooting parameter leads to differences in terms of core masses which
are well below those triggered by a $0.5M_{\odot}$ shift in the total mass
of the star, as seen in Table 1. This is also confirmed by the evolution of
the $4M_{\odot}$ model without overshooting, which achieves lithium production
and by a $3M_{\odot}$ model with $\zeta=0.03$,
which fails to do so. 
We also computed an extensive grid of models
in the range $5.5M_{\odot}\leq M \leq 6.5M_{\odot}$. By adopting $\zeta=0.02$
we found that the maximum value of M wich does not ignite $^{12}C$ in a semi
degenerate regime is $M=6M_{\odot}$, while for $\zeta=0.03$ this limit
is $M=5.8M_{\odot}$.
On the basis of these results we can conclude that the range 
of $3.5M_{\odot}\leq M \leq 6M_{\odot}$
derived in the previous section is well 
extablished within $\sim 0.5M_{\odot}$. 

\section{Conclusions}
In this paper we have presented grids of intermediate mass models of metallicity
appropriate for the LMC AGBs, Z=0.01, in order to reproduce
the observed trend lithium vs luminosity, found by the survey of both 
Magellanic Clouds by Smith et al. (1995). We found that the interval of initial 
masses involved in lithium production is well defined within $0.5M_{\odot}$
even considering all the uncertainties connected with the overshooting
distance and the mass loss rate, and it is
$3.5M_{\odot} \leq M \leq 6M_{\odot}$. More particularly, models
with initial masses $M \geq 5.5M_{\odot}$ display a very peculiar behaviour,
since they produce lithium even before the beginning of the first pulse.

Numerical simulations lead to the conclusion that large mass loss rates,
approaching $10^{-4} M_{\odot}$ yr$^{-1}$, are required to fit the observations,
otherwise we would expect to detect several large luminosity sources
(\Mbol $\simlt -6.5$) with negligible amount of lithium in their envelope,
while the afore mentioned survey shows that practically 
all the AGB sources in the MCs with \Mbol $\leq -6.5$ are lithium rich; 
if we rely on Bl\"ocker's recipe for mass loss, we find that a value 
of the free parameter of $\eta_R=0.01$ is required in our models, 
while values $\eta_R \geq 0.05$ can be disregarded since in these cases
the most luminous Li-rich AGBs would have progenitors masses $M_{ZAMS} 
\simgt 5.5M_{\odot}$. These latter would produce lithium before they have TPs,
so that they would not have s-process enriched envelopes, in 
contrast with the Smith et al. (1995) results. 

We conclude that the most luminous Li-rich AGBs in the LMC 
represent the early AGB phases of the evolution of stars with initial masses
$M\sim 4 - 4.5M_{\odot}$.
Our models of large progenitor mass ($M\sim 6M_{\odot}$) seem to be able 
to give a theoretical explanation of the existence in the LMC of  
AGB sources at \Mbol $= -7.3$ and $-7.6$, 
which are long period, obscured variables (Wood et al 1992).

As a consequence of our calibration of mass loss, massive AGBs
should not contribute significantly to the Lithium enrichment of the
interstellar medium.
\vskip0.2in
\acknowledgements{P.Ventura thanks M.Groenewegen for helpful discussions.
We also thank B.Plez for useful informations on Li-rich AGBs and the
referee A.Zijlstra for a very useful report.}

\end{document}